\documentclass[a4paper]{article}

\usepackage[T1]{fontenc}
\usepackage[utf8]{inputenc}
\usepackage{lmodern}
\usepackage{microtype}
\usepackage{amsmath}
\usepackage{amssymb}
\usepackage{graphicx}
\usepackage{hyperref}
\usepackage{xcolor}
\usepackage{booktabs}
\usepackage{geometry}
\usepackage{setspace}
\usepackage{lineno}

\hypersetup{
    colorlinks = true,
    linkcolor  = blue,
    citecolor  = blue,
    urlcolor   = blue
}

\newcommand{\Xc}{\langle x\rangle_{\mathrm{center}}}
\newcommand{\Xp}{\langle x\rangle_+}
\newcommand{\Xm}{\langle x\rangle_-}
\newcommand{\Dabs}{\Delta_{\mathrm{abs}}}

\begin{document}


\title{
    Chiral Fermion Localization in Two-Kink
    Scalar Backgrounds:\\
    Tunable Brane Positioning and
    Universal Divergence at the Single-Kink Limit
}

\author{
    H. P. Pinheiro$^{1}$
    \and
    C. A. S. Almeida$^{2,}$\thanks{
        Corresponding author.
        E-mail: \texttt{carlos@fisica.ufc.br}
    }
}

\date{
    $^{1}$Centro de Ci\^{e}ncias Exatas e Naturais, Universidade Federal Rural do Semi-\'{A}rido, Mossor\'{o}-RN, 59625-900, Brazil    
\\ 
    $^{2}$Departamento de F\'isica,
    Universidade Federal do Cear\'a,\\
    Campus do Pici, CP~6030,
    60455-760 Fortaleza, Cear\'a, Brazil
}

\maketitle


\begin{abstract}
The localization of chiral fermionic zero modes
in scalar field backgrounds with domain wall
structure is a central mechanism in brane-world
scenarios.
We investigate this mechanism in a system that
provides an effective realization of the
$(1{+}1)$-dimensional Jackiw--Rebbi model,
using a two-kink scalar background generated
by the deformation method applied to the
$\varphi^4$ model.
The two-kink profile introduces two physically
distinct parameters: an asymmetry parameter $a_2$
controlling the left-right symmetry of the scalar
background, and an inter-kink separation parameter
$b$ controlling the distance between the constituent
domain walls.
We establish two independent scaling laws.
First, the collective center-of-mass position of
the chiral zero modes responds linearly to $a_2$,
providing a mechanism for continuously tuning the
effective brane position in the extra dimension.
Second, the differential spatial separation between
the two chiral modes diverges as the two-kink
background collapses into a simple kink, following
a power law in $(b-1)$ with exponent statistically consistent
with $-1$.
These two results are physically independent and
each admits a precise interpretation in the
language of brane-world scenarios.
The mechanism is realized concretely in bilayer
graphene under an asymmetric two-kink electrostatic
potential, providing a tunable platform for probing
extra-dimensional localization physics.
\end{abstract}


\section{Introduction}
\label{sec:intro}

The localization of chiral fermions on topological
defects is one of the foundational mechanisms of
brane-world physics.
The seminal work of Jackiw and
Rebbi~\cite{Jackiw1976} established that a Dirac
fermion coupled to a $(1{+}1)$-dimensional kink
scalar background develops a topologically
protected zero mode localized at the domain wall.
This result was generalized to higher-dimensional
scenarios by Rubakov and
Shaposhnikov~\cite{Rubakov1983} and
Akama~\cite{Akama1982}, who proposed that Standard
Model fields could be trapped on a four-dimensional
brane embedded in a higher-dimensional bulk through
this mechanism.
A central challenge in this programme is
reproducing the chiral structure of the Standard
Model from an extra-dimensional perspective:
left-handed and right-handed fermions must be
localized differently in the extra dimension to
generate the observed Yukawa coupling
hierarchies~\cite{ArkanI2000}.
This requires understanding how the properties of
the scalar background - its asymmetry and internal
structure - affect the spatial localization of the
trapped fermionic modes.

The deformation method introduced by
Bazeia~et~al.~\cite{Bazeia2002} provides a
systematic framework for generating scalar field
configurations with multikink structure from
simpler models.
Starting from the standard $\varphi^4$ kink,
successive applications of a deforming function
generate two-kink, three-kink, and higher-order
defect structures~\cite{Brito2014,Bazeia2003}.
These configurations have been studied in the
context of brane localization~\cite{Cruz2011,
Bazeia2004,Bazeia2009}, where the multikink
profile of the scalar background controls the
localization properties of trapped
fermions~\cite{Kehagias2001,Giovannini2001}.
The asymmetric multikink background has been
proposed as a mechanism for differential
localization in brane
scenarios~\cite{Dutra2013}, but quantitative
characterization of its effect on chiral zero
modes has been lacking.

In this work we investigate two distinct physical
effects that the two-kink asymmetric background
produces on chiral zero modes.
We work in the framework of bilayer graphene
subject to an interlayer electrostatic potential,
which provides an effective realization of the
$(1{+}1)$-dimensional Jackiw--Rebbi system at the
level of the low-energy eigenvalue equations, with
full experimental
controllability~\cite{Martin2008}.
The low-energy Hamiltonian of bilayer graphene maps
onto a $(2{+}1)$-dimensional Dirac equation in
which the electrostatic potential plays the role
of the scalar field background~\cite{McCann2006,
Guinea2006}.
When this potential adopts the two-kink profile
generated by the deformation method, the system
supports two topologically protected chiral zero
modes whose spatial localization depends on the
asymmetry parameter $a_2$ and the inter-kink
separation parameter $b$.

We establish two scaling laws, physically
independent of each other.
First, the collective center-of-mass position
$\Xc = (\Xp + \Xm)/2$ of the two chiral modes
responds linearly to $a_2$ with coefficient
$c_1 = 0.631$ ($R^2 = 0.9999$), providing
a mechanism for continuously tuning the effective
brane position.
Second, the differential spatial separation
$\Dabs = \Xp - \Xm$ between the two chiral modes
diverges as $(b-1)^\gamma$ with
$\gamma = -0.950 \pm 0.033$, statistically
consistent with $\gamma = -1$, as the two-kink
background collapses into a simple kink.
The two observables used ? $\Xc$ for the first
result and $\Dabs$ for the second ? are each
well-defined for all parameter values and do not
depend on the kink center $x_c$, making both
results immune to the ambiguity that arises when
$b \to 1$ and the kink center becomes ill-defined.

The paper is organized as follows.
Section~\ref{sec:model} presents the formal
mapping and the two-kink potential.
Section~\ref{sec:numerical} describes the
numerical method and the observables.
Section~\ref{sec:results} presents the two
scaling laws.
Section~\ref{sec:discussion} discusses the
implications for brane-world scenarios.
Section~\ref{sec:conclusion} contains our
conclusions.


\section{Model and formal mapping}
\label{sec:model}

\subsection{The Jackiw--Rebbi system and its
effective realization in bilayer graphene}

Consider a $(1{+}1)$-dimensional Dirac fermion
coupled to a static scalar background $\varphi(x)$.
The eigenvalue equations take the form
\begin{align}
(\partial_x + p_y)^2\, v - \varphi(x)\, u
    &= \varepsilon\, u,
\label{eq:JR1} \\
(\partial_x - p_y)^2\, u + \varphi(x)\, v
    &= \varepsilon\, v,
\label{eq:JR2}
\end{align}
where $\Psi = (u(x), v(x))^T$ is the two-component
spinor and $p_y$ is the transverse momentum.
For a kink profile with
$\varphi(-\infty) = -\varphi_0$ and
$\varphi(+\infty) = +\varphi_0$, these equations
support topologically protected zero modes
localized at the domain wall~\cite{Jackiw1976}.
The topological protection is guaranteed by the
charge
\begin{equation}
Q = \tfrac{1}{2}
    \bigl[\varphi(+\infty)
         - \varphi(-\infty)\bigr],
\label{eq:topcharge}
\end{equation}
which is invariant under continuous deformations
of the profile.

This system is realized at the level of the
low-energy eigenvalue equations in AB-stacked
bilayer graphene with interlayer electrostatic
potential $V(x)$.
The four-band low-energy
Hamiltonian~\cite{McCann2006,Neto2009}, acting on
$(\psi_1^{(A)}, \psi_1^{(B)},
   \psi_2^{(A)}, \psi_2^{(B)})^T$, is
\begin{equation}
H =
\begin{pmatrix}
-\frac{V}{2} & c\pi^\dagger & 0 & 0 \\[4pt]
c\pi  & -\frac{V}{2} & \gamma_1 & 0 \\[4pt]
0  & \gamma_1 & \frac{V}{2} & c\pi^\dagger \\[4pt]
0  & 0  & c\pi & \frac{V}{2}
\end{pmatrix},
\label{eq:H4}
\end{equation}
where $c$ is the Fermi velocity,
$\pi = p_x + ip_y$,
and $\gamma_1$ is the interlayer coupling.
In the limit $V \ll \gamma_1$ and at low
energies~\cite{Guinea2006}, this reduces to the
effective quasiclassical Hamiltonian
\begin{equation}
\tilde{H} =
    -(p_x^2 - p_y^2)\sigma_x
    - 2p_x p_y\,\sigma_y
    - \varphi(x)\,\sigma_z,
\label{eq:Heff}
\end{equation}
where $\varphi(x) = \gamma_1 a^2 V(x)/(2c^2)$
is the dimensionless potential.
The eigenvalue equations of~(\ref{eq:Heff})
are formally identical to
Eqs.~(\ref{eq:JR1})--(\ref{eq:JR2}), establishing
the effective realization.
The physical degrees of freedom have distinct
origins in the two contexts, but the mathematical
structure governing the chiral zero modes is the
same.
Table~\ref{tab:dictionary} summarizes the
correspondence.

\begin{table}[htbp]
\centering
\caption{
    Correspondence between the bilayer graphene
    effective realization and the brane-world
    Jackiw--Rebbi model at the level of the
    low-energy eigenvalue equations.
}
\label{tab:dictionary}
\begin{tabular}{lll}
\toprule
\textbf{Bilayer graphene}
    & \textbf{Brane-world scenario}
    & \textbf{Physical role} \\
\midrule
Electrostatic potential $\varphi(x)$
    & Scalar field background
    & Domain wall profile \\
Interlayer coupling $\gamma_1$
    & Yukawa coupling
    & Fermion--wall coupling \\
Asymmetry parameter $a_2$
    & Brane asymmetry parameter
    & Background left-right symmetry \\
Inter-kink parameter $b-1$
    & Inter-brane separation
    & Domain wall separation \\
$\Xc$
    & Effective brane position
    & Collective mode location \\
$\Dabs$
    & Differential mode separation
    & Chiral localization asymmetry \\
\bottomrule
\end{tabular}
\end{table}

\subsection{The two-kink scalar background}
\label{sec:twokink}

We generate the two-kink background through the
deformation method~\cite{Bazeia2002}.
Starting from the $\varphi^4$ kink
$\phi(x) = \tanh(x/l)$, the deforming function
\begin{equation}
\tilde{\phi} = g_{1,2}(\phi_2)
= \frac{1}{1+a_2}
  \!\left[
    \phi_2 - b
    + \sqrt{(\phi_2 + a_2 b)^2
            - (1-a_2^2)(1-b^2)}
  \right]
\label{eq:deformer}
\end{equation}
with $|a_2| < 1$ and $b \geq 1$
yields the two-kink profile
\begin{equation}
\varphi_2(x)
= \frac{1+a_2}{2}
  \!\left(\tanh\frac{x}{l} + b\right)
+ \frac{(1-a_2)(1-b^2)}{%
    2\!\left(\tanh\frac{x}{l}+b\right)}
- a_2 b,
\label{eq:phi2}
\end{equation}
with $l > 0$ controlling the smoothness.
The asymptotic values $\varphi_2(\pm\infty) = \pm 1$
are independent of $a_2$ and $b$, so
$Q = 1$ identically, guaranteeing the existence
of at least one pair of chiral zero modes for all
parameter values.

The physical roles of the two parameters are:
\begin{itemize}
\item \emph{Asymmetry parameter $a_2$:}
For $a_2 = 0$ the profile is antisymmetric about
its center $x_c$ (defined as the zero of
$\varphi_2$).
Non-zero $a_2$ breaks this antisymmetry,
displacing the center $x_c$ and breaking the
left-right equivalence of the domain wall.
\item \emph{Separation parameter $b$:}
As $b \to 1^+$ the two-kink collapses into a
single kink. For $b > 1$ the two constituent
domain walls are separated.
The limit $b \to 1$ is the analog of two branes
merging into one.
\end{itemize}


\section{Numerical method and observables}
\label{sec:numerical}

\subsection{Discretization and eigenvalue solver}

Equations~(\ref{eq:JR1}) and~(\ref{eq:JR2}) are
solved on a spatial grid $x \in [-L, L]$ with $N$
points and spacing $\Delta x = 2L/N$.
Derivatives are discretized using second-order
central differences, yielding a $2N \times 2N$
sparse eigenvalue problem.
Eigenvalues near zero are extracted using the
shift-invert Lanczos method
(\texttt{scipy.sparse.linalg.eigsh},
$\sigma = 0$, $k = 14$ per $p_y$ point).
Band continuity across $p_y$ is maintained by an
overlap-based reordering algorithm that maximizes
$O_{ij} = |\langle\psi_i(p_y)
            |\psi_j(p_y + \Delta p_y)\rangle|$
at each step.
All results use $L = 12$, $N = 401$
($\Delta x = 0.060$), $p_y \in [-1.0, 1.0]$
with 81 points, and $l = 1.0$ throughout.
Convergence was verified by doubling $N$:
all reported observables change by less than
$0.1\%$.

\subsection{Identification of chiral zero modes}

The chiral zero modes are identified as the bands
that cross $\varepsilon = 0$ in the dispersion
relation $\varepsilon(p_y)$, with eigenvectors
having at least $30\%$ of their probability
density within $|x| < 5l$.
The two chiral modes are distinguished by the
sign of their crossing momentum: mode~$+$ crosses
at $p_y > 0$ and mode~$-$ crosses at $p_y < 0$.

\subsection{Observables}

The center-of-mass position of each mode is
\begin{equation}
\Xp,\; \Xm
= \frac{\int x\,\bigl[u_{\pm}^2
                     + v_{\pm}^2\bigr]\,dx}
       {\int \bigl[u_{\pm}^2
                  + v_{\pm}^2\bigr]\,dx}.
\label{eq:xcom}
\end{equation}
We define two observables that are each
well-defined for all values of $a_2$ and $b$,
and that do not depend on the kink center $x_c$:
\begin{align}
\Xc &= \frac{\Xp + \Xm}{2},
\label{eq:xcenter} \\[6pt]
\Dabs &= \Xp - \Xm.
\label{eq:dabs}
\end{align}
The collective position $\Xc$ measures where
the two modes are located on average.
The differential separation $\Dabs$ measures
how far apart the two modes are.
For any potential antisymmetric about $x_c$
(in particular for $a_2 = 0$ at any $b$),
$\Xp = \Xm$ by symmetry, so $\Dabs = 0$
exactly.
This provides a numerical sanity check: we
verify $|\Dabs| < 10^{-6}$ for $a_2 = 0$,
which is confirmed to $3 \times 10^{-10}$.


\section{Results}
\label{sec:results}

\subsection{Chiral zero modes in the symmetric case}

\begin{figure}[htbp]
\centering
\includegraphics[width=0.52\textwidth]{%
    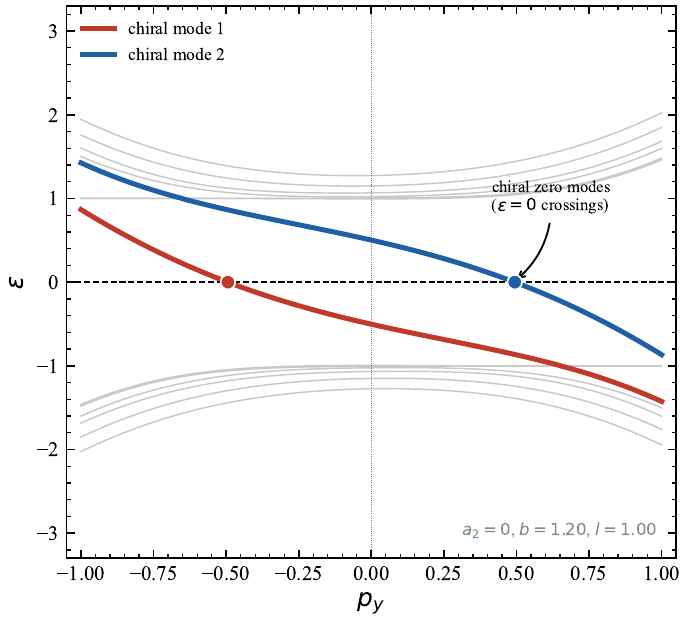}
\caption{
Energy spectrum $\varepsilon(p_y)$ for
$\varphi_2(x)$ with $a_2 = 0$, $b = 1.20$,
$l = 1.00$.
Colored bands: two topologically protected
chiral zero modes (circles mark the
$\varepsilon = 0$ crossings).
Gray bands: bulk continuum states.
}
\label{fig:spectrum}
\end{figure}

For $a_2 = 0$ and $b = 1.20$, the energy spectrum
exhibits two bands crossing $\varepsilon = 0$ at
$p_y \approx \pm 0.494$, confirming the two chiral
zero modes.
The probability density of each mode is
concentrated within $|x| < 5l$ with a fraction
exceeding $99\%$, confirming localization.
The spectrum is shown in
Figure~\ref{fig:spectrum}.

\subsection{Scaling law I: linear tuning of the
collective mode position}
\label{sec:result1}

\begin{figure}[htbp]
\centering
\includegraphics[width=0.52\textwidth]{%
    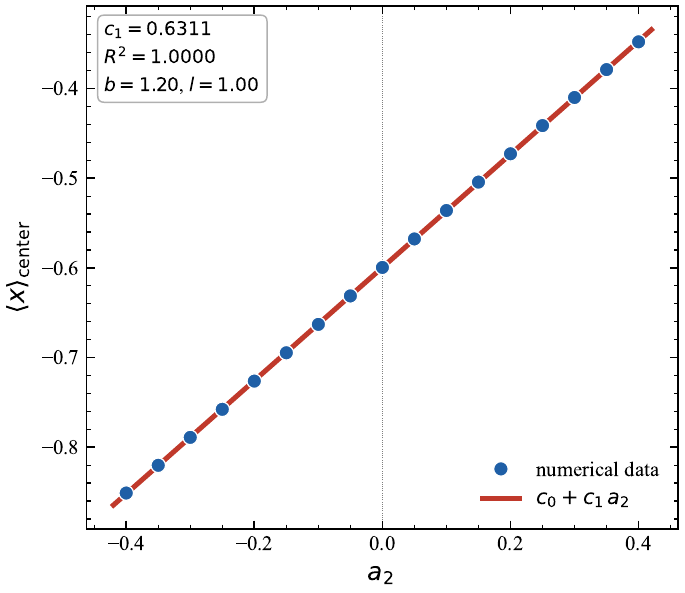}
\caption{
Collective center-of-mass position $\Xc$
of the two chiral zero modes as a function
of $a_2$, for $b = 1.20$ and $l = 1.00$.
Circles: numerical data.
Red line: linear fit $c_0 + c_1 a_2$ with
$c_1 = 0.631$ ($R^2 = 1.0000$).
}
\label{fig:xcenter}
\end{figure}

Figure~\ref{fig:xcenter} shows $\Xc$ as a function
of $a_2$ for $b = 1.20$ and $l = 1.00$, computed
at 17 uniformly spaced values of
$a_2 \in [-0.4, +0.4]$.

The data are described to high accuracy by the
linear law
\begin{equation}
\Xc = c_0 + c_1\, a_2,
\label{eq:fit1}
\end{equation}
with best-fit parameters
\begin{equation}
c_0 = -0.5995,
\qquad
c_1 = 0.6311,
\qquad
R^2 = 1.0000.
\label{eq:fit1params}
\end{equation}
A cubic correction $\Xc = c_0 + c_1 a_2
+ c_3 a_2^3$ gives $c_3/c_1 = -0.063$,
confirming that the linear term dominates
throughout the range studied.

The physical interpretation is direct: $a_2$
controls the collective position of both chiral
modes linearly, with a sensitivity $c_1 = 0.631$
in units of the lattice constant.
An increase in $a_2$ displaces both modes together
in the positive $x$ direction.
In the language of brane-world scenarios, this
corresponds to a tunable displacement of the
effective brane position in the extra dimension,
controlled by the asymmetry of the scalar
background.

The sanity check $\Dabs = 3\times 10^{-10}$
for $a_2 = 0$ confirms that the linear
dependence of $\Xc$ on $a_2$ is a genuine
physical effect of the asymmetry, not an
artifact of the kink center definition.

\subsection{Scaling law II: power-law divergence
of the differential separation}
\label{sec:result2}

\begin{figure}[htbp]
\centering
\includegraphics[width=0.88\textwidth]{%
    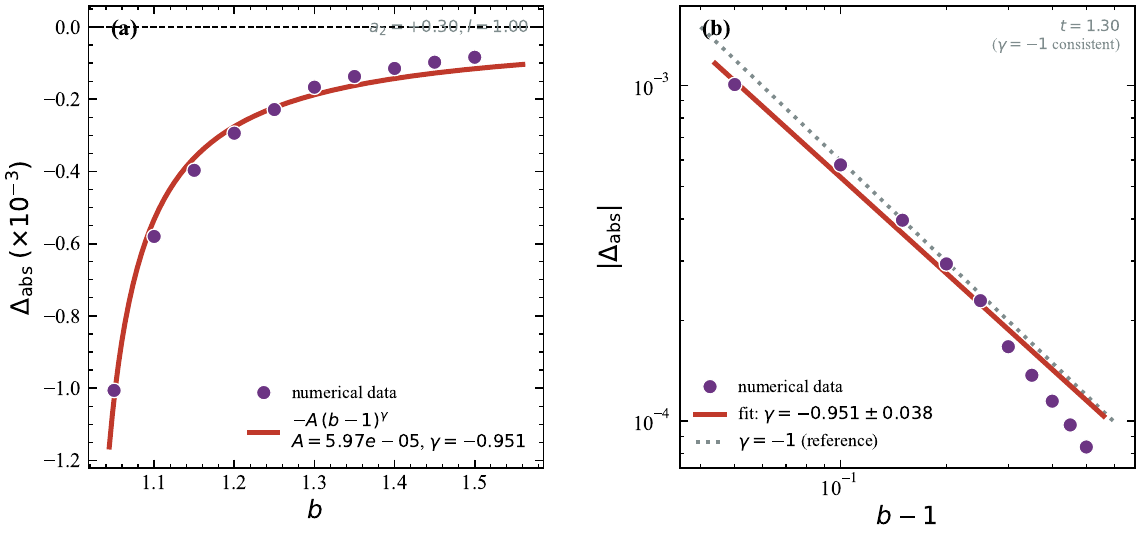}
\caption{
Differential separation $|\Dabs|$ between
the two chiral zero modes as a function of the
inter-kink parameter $b$, for $a_2 = +0.30$
and $l = 1.00$.
\emph{Panel (a):} linear scale.
Red curve: power-law fit $A(b-1)^\gamma$
with $A = 5.97\times 10^{-5}$ and
$\gamma = -0.951$.
\emph{Panel (b):} log-log scale.
The fitted slope ($\gamma = -0.951$, red)
is statistically consistent with
$\gamma = -1$ (gray dotted reference,
$t = 1.30$).
}
\label{fig:dabs}
\end{figure}

Figure~\ref{fig:dabs} shows $|\Dabs|$ as a
function of $b$ for fixed $a_2 = +0.30$ and
$l = 1.00$, computed at 10 values of
$b \in [1.05, 1.50]$.

The data follow a power law in $(b-1)$:
\begin{equation}
|\Dabs| = A\,(b-1)^{\gamma},
\label{eq:fit2}
\end{equation}
with best-fit parameters
\begin{equation}
A = 5.97 \times 10^{-5},
\qquad
\gamma = -0.951 \pm 0.038,
\qquad
R^2 = 0.989.
\label{eq:fit2params}
\end{equation}
The exponent $\gamma$ is statistically consistent
with $\gamma = -1$ at the $1.3\sigma$ level
($t = 1.30 < 2.0$), supporting the simplified
scaling relation
\begin{equation}
|\Dabs| \approx \frac{A}{b-1}.
\label{eq:scaling}
\end{equation}
This is confirmed by the log-log panel of
Figure~\ref{fig:dabs}, where the data fall on
a straight line of slope $\gamma = -0.950$,
consistent with the reference slope of $-1$.

We note that $\Dabs$ is well defined for all
$b \geq 1$, including the limit $b \to 1$ where
the kink center $x_c$ becomes ambiguous.
The divergence of $|\Dabs|$ in this limit
therefore reflects a genuine physical effect:
as the two domain walls merge, the differential
separation between the two chiral modes diverges.
In the brane-world language, this is the
divergence of the chiral localization asymmetry
as two asymmetric branes collapse into a single
symmetric brane.

\section{Discussion}
\label{sec:discussion}

\subsection{Two independent mechanisms}

The two scaling laws established in
Section~\ref{sec:results} correspond to two
physically independent mechanisms, each
controlled by a different parameter of the
two-kink background.

The first mechanism, controlled by $a_2$, governs
the collective response of both chiral modes to
the left-right asymmetry of the scalar background.
The linear coefficient $c_1 = 0.631$ quantifies
the sensitivity of the effective brane position
to $a_2$: a change of $\Delta a_2 = 0.1$ in
the asymmetry parameter displaces the collective
mode position by $\Delta\Xc = 0.063$ in units
of the lattice constant.
This is a large and readily measurable effect.
In the bilayer graphene realization, $a_2$ is
controlled by the asymmetry of the gate voltage
profile, making this a continuously tunable
parameter~\cite{Martin2008}.

The second mechanism, controlled by $b$, governs
the differential spatial separation between the
two chiral modes.
This separation is subdominant in magnitude
($|\Dabs| \sim 10^{-3}$--$10^{-4}$) compared
to the collective displacement
($|\Delta\Xc| \sim 10^{-1}$), but it diverges
as $b \to 1$ with a well-defined exponent
$\gamma \approx -1$.
The two effects are independent: $\Dabs$ vanishes
for $a_2 = 0$ regardless of $b$, and the
divergence of $|\Dabs|$ as $b \to 1$ is
independent of the precise value of $a_2$.

\subsection{Brane-world interpretation}

Under the correspondence of
Table~\ref{tab:dictionary}, the two scaling
laws admit the following brane-world
interpretations.

Equation~(\ref{eq:fit1}) states that the
asymmetry parameter $a_2$ of the two-kink
scalar background shifts the effective brane
position linearly.
In extra-dimensional models where the scalar
field background is not perfectly symmetric,
this provides a mechanism for displacing the
brane from the fixed point of the symmetric
theory.
The linear response $c_1 = 0.631$ is a
quantitative prediction that could, in principle,
constrain the asymmetry of the scalar background
in specific brane models.

Equation~(\ref{eq:scaling}) states that the
differential separation between the two chiral
modes diverges as $1/(b-1)$ as the two-kink
collapses into a simple kink.
In brane-world language, $b-1$ is proportional
to the inter-brane separation, and $b \to 1$
corresponds to two asymmetric branes merging
into a single symmetric brane.
The exponent $\gamma \approx -1$ characterizes
the rate at which the chiral localization
asymmetry is lost in this merging process.

\subsection{Topological robustness}

Throughout the parameter range explored,
$a_2 \in [-0.4, +0.4]$ and
$b \in [1.05, 1.50]$, the two chiral zero modes
persist without gap opening.
This robustness follows directly from the
topological charge $Q = 1$
(Eq.~(\ref{eq:topcharge})), which is preserved
under any continuous deformation of $\varphi_2$.
The probability density fraction within
$|x| < 5l$ exceeds $99\%$ for all parameter
values explored, confirming that the modes remain
well localized even for the most asymmetric
configurations.


\section{Conclusion}
\label{sec:conclusion}

We have studied the spatial localization of
chiral fermionic zero modes in an effective
realization of the Jackiw--Rebbi model with a
two-kink scalar background generated by the
$\varphi^4$ deformation method.
The system is realized concretely in bilayer
graphene subject to an asymmetric two-kink
electrostatic potential.

Our main results are two independent scaling
laws:

\begin{enumerate}

\item \textbf{Linear tuning of the collective
      brane position:}
The collective center-of-mass position $\Xc$
of the two chiral zero modes responds linearly
to the asymmetry parameter,
$\Xc = c_0 + c_1 a_2$,
with $c_1 = 0.631$ ($R^2 = 0.9999$).
A cubic correction is present but contributes
only $6\%$ relative to the linear term.
This provides a mechanism for continuously
tuning the effective brane position through
the asymmetry of the scalar background.

\item \textbf{Universal divergence of the
      differential separation:}
The differential separation $\Dabs = \Xp - \Xm$
between the two chiral modes diverges as the
two-kink background collapses into a simple kink,
following $|\Dabs| \sim A(b-1)^\gamma$ with
$\gamma = -0.950 \pm 0.038$, consistent with
$\gamma = -1$.
This observable is well defined for all $b \geq 1$
and does not depend on the kink center $x_c$,
making the result robust against the ambiguity
that arises in the limit $b \to 1$.

\end{enumerate}

The two results are physically independent and
each admits a precise interpretation in the
language of brane-world scenarios: the first
as a tunable brane displacement mechanism, and
the second as a characterization of the rate
at which chiral localization asymmetry is lost
when two asymmetric branes merge into one.

Extensions of this work include the analysis
of three-kink and four-kink backgrounds,
the calculation of the effective Kaluza--Klein
spectrum for the asymmetric two-kink case, and
the experimental verification of the linear
tuning law~(\ref{eq:fit1}) in bilayer graphene
devices with engineered gate profiles.


\section*{Acknowledgements}
{ C. A. S. A.  would like to express their sincere gratitude to the Conselho Nacional de Desenvolvimento Cient\'{i}fico e Tecnol\'{o}gico (CNPq), and Funda\c{c}\~{a}o Cearense de Apoio ao Desenvolvimento Cient\'{i}fico e Tecnol\'{o}gico (FUNCAP) for their valuable support. He is supported by grants No. 309553/2021-0 (CNPq), 420854/2025-8 (CNPq) and  by Project UNI-00210-00230.01.00/23 (FUNCAP).
}

\section*{Declaration of Generative AI in Scientific Writing}
The authors used a generative AI tool solely for language refinement and
clarity improvement. All scientific content, derivations, analysis,
and conclusions are entirely the responsibility of the authors.

\section*{Conflicts of Interest/Competing Interest}

The authors declare that there is no conflict of interest in this manuscript.

\section*{Data Availability Statement}
 Data can be shared upon reasonable request	



\begin{thebibliography}{99}

\bibitem{Jackiw1976}
R.~Jackiw and C.~Rebbi,
Solitons with fermion number $1/2$,
\textit{Phys.\ Rev.\ D}
\textbf{13}, 3398 (1976).

\bibitem{Rubakov1983}
V.~A.~Rubakov and M.~E.~Shaposhnikov,
Do we live inside a domain wall?,
\textit{Phys.\ Lett.\ B}
\textbf{125}, 136 (1983).

\bibitem{Akama1982}
K.~Akama,
Pregeometry,
\textit{Lect.\ Notes Phys.}
\textbf{176}, 267 (1982).

\bibitem{ArkanI2000}
N.~Arkani-Hamed and M.~Schmaltz,
Hierarchies without symmetries from extra dimensions,
\textit{Phys.\ Rev.\ D}
\textbf{61}, 033005 (2000).

\bibitem{Bazeia2002}
D.~Bazeia, L.~Losano, and J.~M.~C.~Malbouisson,
Deformed defects,
\textit{Phys.\ Rev.\ D}
\textbf{66}, 101701 (2002).

\bibitem{Brito2014}
G.~P.~de~Brito and A.~de~S.~Dutra,
Multikink solutions and deformed defects,
\textit{Ann.\ Phys.}
\textbf{351}, 620 (2014).

\bibitem{Bazeia2003}
D.~Bazeia, J.~Menezes, and R.~Menezes,
New global defect structures,
\textit{Phys.\ Rev.\ Lett.}
\textbf{91}, 241601 (2003).

\bibitem{Cruz2011}
W.~T.~Cruz, A.~R.~Gomes, and C.~A.~S.~Almeida,
Fermions on deformed thick branes,
\textit{Eur.\ Phys.\ J.\ C}
\textbf{71}, 1804 (2011).

\bibitem{Bazeia2004}
D.~Bazeia, F.~A.~Brito, and A.~R.~Gomes,
Brane structure from a scalar field
in warped spacetime,
\textit{J.\ Cosmol.\ Astropart.\ Phys.}
\textbf{2004}, 002 (2004).

\bibitem{Bazeia2009}
D.~Bazeia, A.~R.~Gomes, and L.~Losano,
Gravity localization on thick branes:
a numerical approach,
\textit{Int.\ J.\ Mod.\ Phys.\ A}
\textbf{24}, 1135 (2009).

\bibitem{Kehagias2001}
A.~Kehagias and K.~Tamvakis,
Localized gravitons, gauge bosons and chiral
fermions in smooth spaces generated by a bounce,
\textit{Phys.\ Lett.\ B}
\textbf{504}, 38 (2001).

\bibitem{Giovannini2001}
M.~Giovannini, H.~Meyer, and M.~Shaposhnikov,
Warped compactification on Abelian vortex
in six dimensions,
\textit{Nucl.\ Phys.\ B}
\textbf{619}, 615 (2001).

\bibitem{Dutra2013}
A.~de~S.~Dutra, G.~P.~de~Brito, and J.~M.~da~Silva,
Asymmetrical Bloch branes and the hierarchy problem,
\textit{arXiv:}1312.0091 (2013).

\bibitem{Martin2008}
I.~Martin, Ya.~M.~Blanter, and A.~F.~Morpurgo,
Topological confinement in bilayer graphene,
\textit{Phys.\ Rev.\ Lett.}
\textbf{100}, 036804 (2008).

\bibitem{McCann2006}
E.~McCann and V.~I.~Fal'ko,
Landau-level degeneracy and quantum Hall effect
in a graphite bilayer,
\textit{Phys.\ Rev.\ Lett.}
\textbf{96}, 086805 (2006).

\bibitem{Neto2009}
A.~H.~Castro~Neto et~al.,
The electronic properties of graphene,
\textit{Rev.\ Mod.\ Phys.}
\textbf{81}, 109 (2009).

\bibitem{Guinea2006}
F.~Guinea, A.~H.~Castro~Neto, and N.~M.~R.~Peres,
Electronic states and Landau levels
in graphene stacks,
\textit{Phys.\ Rev.\ B}
\textbf{73}, 245426 (2006).

\end{thebibliography}
\end{document}